\documentclass{aastex}
\usepackage[onecolumn]{emulateapj5}
\usepackage{onecolfloat5}
\usepackage{apjfonts}
\usepackage{amsmath}

\def\plotone#1{\centering \leavevmode
\epsfxsize= 1.0\columnwidth \epsfbox{#1}}

\def\gsim{\;\rlap{\lower 2.5pt
 \hbox{$\sim$}}\raise 1.5pt\hbox{$>$}\;}
\def\lsim{\;\rlap{\lower 2.5pt
   \hbox{$\sim$}}\raise 1.5pt\hbox{$<$}\;}

\newcommand{\be}{\begin{equation}}
\newcommand{\beq}{\begin{equation}}
\newcommand{\ba}{\begin{eqnarray}}
\newcommand{\ee}{\end{equation}}
\newcommand{\eeq}{\end{equation}}
\newcommand{\ea}{\end{eqnarray}}

\usepackage{epsfig}

\begin{document}
\twocolumn[
\title{Constraints on the Abundance of Highly Ionized Proto-Cluster Regions\\from the Absence of Large Voids in the Lyman Alpha Forest}

\author{Cien Shang}
\affil{Department of Physics, Columbia University, 550 West 120th
Street, New York, NY 10027, USA; cien@phys.columbia.edu}

\author{Arlin Crotts, Zolt\'an Haiman}
\affil{Department of Astronomy, Columbia University, 550 West
120th Street, New York, NY 10027, USA;
(arlin,zoltan)@astro.columbia.edu}

\begin{abstract}
Energetic feedback processes during the formation of galaxy clusters
may have heated and ionized a large fraction of the intergalactic gas
in proto--cluster regions. When such a highly ionized hot
``super--bubble'' falls along the sightline to a background quasar, it
would be seen as a large void, with little or no absorption, in the
Lyman--$\alpha$ forest.  We examine the spectra of 137 quasars in the
Sloan Digital Sky Survey, to search for such voids, and find no clear
evidence of their existence. The size distribution of voids in the
range 5\AA\ $\lsim \Delta\lambda\lsim 70$\AA\ (corresponding to
physical sizes of $3h^{-1}\lsim R \lsim 35h^{-1}$ comoving Mpc) is
consistent with the standard model for the Lyman $\alpha$ forest
without additional hot bubbles.  We adapt a physical model for HII
bubble growth during cosmological reionization (Furlanetto,
Zaldarriaga \& Hernquist 2004), to describe the expected
size--distribution of hot super--bubbles at $z\sim 3$.  This model
incorporates the conjoining of bubbles around individual neighboring
galaxies. Using the non--detection of voids, we find that models in
which the volume filling factor of hot bubbles exceeds $\sim 20$
percent at $z\sim 3$ can be ruled out, primarily because they
overproduce the number of large ($40-50$\AA) voids.  We conclude that
any pre--heating mechanism that explains galaxy cluster observations
must avoid heating the low--density gas in the proto--cluster regions,
either by operating relatively recently ($z\lsim 3$) or by depositing
entropy in the high--density regions.
\end{abstract}
]

\section{Introduction}
\label{sec:introduction}

Observations of galaxy clusters suggest that feedback played an
important role during their formation.  The simplest models for galaxy
clusters neglect feedback and assume the gravitational collapse of a
dark matter halo, accompanied by gas infall.  These models fail to
reproduce either the observed scaling relations between bulk
characteristics, or the structural properties of individual clusters
(e.g. Bialek et al. 2001; Voit et al. 2002). For example, the
self-similar gas distribution expected in this model predicts the
relation between X--ray luminosity and temperature, $L_X\propto T^{2}$
(Kaiser 1986), whereas observations find a steeper relation, closer to
$L\propto T^{\sim 3}$ (e.g., David et al. 1993; Arnaud \& Evrard 1999;
Helsdon \& Ponman 2000).  A compelling suggestion to explain the
discrepancy is that the intra--cluster gas has been pre--heated, i.e.,
raised to a higher adiabat, at an early stage in the formation of the
cluster.  The resulting so--called ``entropy floor'' would then
preferentially affect low--mass clusters. Indeed, the observed $L_X-T$
and related scaling relations are well reproduced in models that
simply endow the gas by an extra entropy, of order $\sim 100~{\rm keV
cm^{-2}}$, before its collapse (Voit \& Bryan 2001, Bialek et
al. 2001; Voit, Bryan, Balogh \& Bower 2002; McCarthy et al. 2003a,b).
The physical mechanism responsible for the pre--heating could be
supernova--driven galactic winds, or the radiation output of active
galactic nuclei (AGN).

The simplest form of this pre--heating model does not appear to
provide an acceptable fit to the detailed cluster profiles (Pratt et
al. 2005, 2006; Younger \& Bryan 2007).  Nevertheless, the broader
idea, namely that energetic processes strongly influenced at least
parts of the intergalactic medium (IGM), corresponding to
proto--cluster regions, at early times, remains viable.  Indeed, there
is considerable empirical support for this broader picture.  The
global star--formation rate, inferred from galaxies discovered at
redshift $z\sim 3$, such as the so--called Lyman-break galaxies
(LBGs), appears significantly higher than the star formation rate in
the local universe.  Energetic "superwinds" from LBGs at $z\sim 3$
have been inferred directly from their UV spectra, showing
several--hundred km/s offsets between stellar and interstellar lines
(Heckman et al. 2000; Pettini et al 2001).  Similar winds are known to
accompany nearby star--bursts (e.g. Heckman et al. 1990) and such
winds would be natural candidates for large--scale feedback at earlier
times.

Indeed, various studies have suggested that winds from galaxies can
affect not only the galaxy itself, but also the surrounding IGM out to
a distance approaching $\sim 1$ (comoving) $h^{-1}$ Mpc, which may
affect global Lyman $\alpha$ absorption statistics (e.g. Fang et
al. 2005; Kollmeier et al. 2006; Desjacques et al. 2006).  Recent
works have focused on interpreting observations of Lyman $\alpha$
absorption statistics in quasar spectra with sightlines passing near
LBGs. The observations possibly indicate a reduced level of absorption
within $\sim 1 h^{-1}$ Mpc of LBGs (Adelberger et al. 2003; 2005),
which may be attributable to the impact of these galaxies on the
ambient IGM (but see Desjacques et al. 2006).  In the context of the
LBGs, several groups have used numerical simulations to study the
metal--enrichment of the IGM by galactic outflows, and the
corresponding impact on the global Lyman $\alpha$ absorption
statistics (Theuns et al. 2002; Bruscoli et al. 2003; McDonald et
al. 2005).

The present study is motivated by the related suggestion of Theuns et
al. (2001), that the preheating of large proto--cluster regions may
leave a direct imprint on the global Lyman $\alpha$ forest absorption
statistics.  Irrespective of the physical mechanism, the pre--heating
would likely ionize the hydrogen in the proto--cluster region, and the
resulting hot bubble would be optically thin in Lyman line absorption.
Theuns et al. (2001) proposed that if the suggested entropy level does
exist, the highly ionized proto--cluster regions could produce large
voids: stretches of wavelength as long as $\sim 20$\AA\ with little or
no absorption.

Such hot proto--cluster bubbles could be an order of magnitude larger
(in linear size) than the $\lsim 1 h^{-1}$ Mpc ionized bubbles that
may envelope individual LBGs. These large proto--cluster bubbles may,
of course, correspond to a clustered group of bubbles around LBGs.  On
the other hand, they could be produced by the collective effect of a
group of galaxies that are much smaller and/or formed earlier than the
known population of LBGs. In this case, the large voids may be more
readily identified in studying the global Lyman $\alpha$ forest
statistics.

Historically, a few authors have searched for large voids in the
Lyman--$\alpha$ forest (Atwood et al. 1985; Crotts 1987; Ostriker,
Bajtlik \& Duncan 1988; Duncan, Ostriker \& Bajtlik 1989; Dobrzycki \&
Bechtold 1991).  In the standard model for the Lyman--$\alpha$ forest,
the absorption lines are produced by fluctuations in the density
field. Observed statistics, such as the column density distribution
and evolution, and the spatial distribution of the absorbers, are
consistent with a model in which the gas traces the primordial
dark--matter fluctuations, and is kept photoionized by a uniform
metagalactic radiation (Miralda--Escud\'e et al. 1996; Hui \& Gnedin
1997).  On large scales ($\gsim 10h^{-1}$ Mpc), the Lyman--$\alpha$
absorbers are essentially randomly distributed in space, and their
incidence rate statistics in quasar spectra are described by a Poisson
distribution.  The studies listed above have identified only a handful
of candidates for large voids that were discrepant with a Poisson
distribution (and not associated with the proximity effect of the
background quasar itself), but none of these have been confirmed at
high statistical significance.

In this paper, we perform a new search for large voids in the
Lyman--$\alpha$ forest.  Our analysis differs from existing studies in
two important ways.  First, we use a larger sample of quasar spectra,
available from the Sloan Digital Sky Survey (SDSS). Second, while we
adopt the same null--hypothesis as previous works (i.e. a Poisson
distribution for the absorbers), we use a new physical model for the
bubble distribution that tracks the conjoining of bubbles around
individual galaxies.  These ``mergers'' between bubbles are important
when their volume--filling factor rises above a few percent: galaxies
are clustered in space, and a single large void will typically contain
many galaxies.  As a result, mergers are a way to produce larger,
possibly detectable voids.


The rest of this paper is organized as follows.
In \S~\ref{sec:model}, we explain how we model the mass function of
highly ionized regions.
In \S~\ref{sec:data}, we briefly describe the observational data that
we used.
In \S~\ref{sec:stat}, we introduce our approach of statistically
comparing the predicted and observed void distribution in
Lyman--$\alpha$ forest.
In \S~\ref{sec:results}, we present our main results.
In \S~\ref{sec:discussion}, we discuss the limitations of our
approach, as well as possible future improvements.
In \S~\ref{sec:conclude}, we briefly summarize our conclusions and
the implications of this work.
Throughout this paper, we adopt a spatially flat universe dominated by
a cosmological constant and cold dark matter (CDM), with the following
set of cosmological parameters: $\Omega_m=0.3$,
$\Omega_{\Lambda}=0.7$, $\sigma_{8}=0.9$ and $H_0=70~{\rm
  km~s^{-1}~Mpc^{-1}}$.  These values are consistent with measurements
by the {\it WMAP} experiment (Spergel et al. 2003; 2007; we include a
discussion of the sensitivity of our results to the choice of
$\sigma_8$ below).


\vspace{3\baselineskip}
\section{Modeling the distribution of the highly ionized regions}
\label{sec:model}

Our main task is to model the abundance and size distribution of
highly ionized proto--cluster regions.  In the back--of--the envelope
style estimate in Theuns et al. (2001), a proto-cluster that later
develops into a cluster of mass $M$ was treated as a uniform sphere
containing the same amount $M$ of fully ionized gas at some fixed
overdensity $\delta$ relative to the cosmic mean gas density at
redshift $z$.  Here $z$ and $\delta$ are both free parameters, the
relevant values of which would need to be estimated from some further
modeling. Assuming that pre--heating operates at redshift $z\sim 3$
and that the mean overdensity in the proto--cluster region is
$\delta\sim 1$, they calculated the proto--cluster size distribution
from the known mass function of galaxy clusters. They concluded that
the typical size of a void at this redshift should be a few$\times
10$\AA, which would appear prominent in high--redshift ($z\geq 3$)
quasar spectra.  Based on the local abundance of massive clusters,
they estimated that there should be approximately one such void per
unit redshift.

An obvious refinement of this simple estimate is to make a connection
between redshift and overdensity by using the spherical collapse
model. Given the current overdensity of clusters and their collapse
redshift, the cluster's expected overdensity at some higher redshift
can be obtained directly. This approach would eliminate one free
parameter, but would still miss an important ingredient of cluster
formation: mergers. In the hierarchical structure formation scenario,
proto--clusters are more likely to be made up of many smaller clumps
that would later merge together. Star formation and AGN activity in
these clumps (which could represent an individual galaxy, or a small
group of highly clustered galaxies) could then ionize gas in their
vicinity. Hereafter we will refer to the area ionized around a single
progenitor clump as a "hot bubble". Several hot bubbles, initially
generated independently in different collapsed regions, may overlap
with each other, and form a larger ``super--bubble''. In order to
compute the distribution of voids in the Lyman--$\alpha$ forest, we
first need to get the mass function of these
super--bubbles. 

Furlanetto et al. (2004) have studied an analogous problem for the
mergers of ionized bubbles at higher redshifts, in the context of
cosmological reionization.  Here we adopt their formalism, and apply
it to the super--bubbles at lower redshift.  In the context of
reionization, the formalism has been compared to numerical simulations
of the growth of ionized bubbles, and was found to accurately
reproduce the size--distribution and large--scale clustering
properties of ionized bubbles (Zahn et al. 2007).  We caution,
however, that a similar test against simulations has not yet been
performed at lower redshifts (see discussion below).  The formalism is
based on the simple assumption that a collapsed DM halo can ionize a
region whose mass is proportional to the halo's own mass.  The
effective proportionality coefficient, denoted as $\zeta$,
\begin{eqnarray}
m_{ion}=\zeta m_{col} 
\label{eq:assumption}
\end{eqnarray}
depends, in the context of reionization, on the efficiency of ionizing
photon production, escape fraction of these photons from the host
galaxy, the star-formation efficiency, and the mean number of
recombinations.  In our case, we define an analogous coefficient
between the mass of a bubble and the mass of a halo,
\begin{eqnarray}
m_{bubble}=\zeta m_{col}.
\label{eq:mbubble}
\end{eqnarray}
The value of this coefficient should depend on the velocity,
temperature, and typical age of galactic winds, or, alternatively, on
similar parameters for the typical AGN outflows.  However, this
simplified description could plausibly describe other scenarios, as
well (e.g. a simple photo--ionization proximity effect).

The condition $\zeta>1$ has to be satisfied in order for the winds to
propagate outside the DM halos and generate hot bubbles in the IGM. In
this case, there is a chance that different bubbles can overlap and
unite into a larger super--bubble. The statistics of the super--bubble
size distribution is, in general, then driven by this overlap, which,
in turn, is governed by the large--scale density fluctuations. In
order to avoid modeling the complex process of overlap, Furlanetto et
al. propose to utilize the following relation, which must be satisfied
for every super--bubble:
\begin{eqnarray}
f_{coll}\geqslant\zeta^{-1}
\label{eq:condition}
\end{eqnarray}
where $f_{coll}$ is the collapsed fraction (the ratio of mass residing
in collapsed halos to the total mass inside the super--bubble), and is
determined using the extended Press--Schechter model. In this
approach, $f_{coll}$ depends on the mean linear overdensity
$\delta_{m}$ inside the super--bubble. The excursion--set formalism
can be used to find the largest region surrounding an arbitrary point
in space, where the above relation is satisfied. The final result for
the mass function of super--bubbles is
\begin{eqnarray}
m\frac{\mathrm{d} n}{\mathrm{d} m}=\sqrt{\frac{2}{\pi}}\frac{\bar{\rho}}{m}\left|\frac{\mathrm{d\ ln}\sigma}{\mathrm{d\ ln}m}\right|
\frac{B_{0}}{\sigma(m)}\exp\left[-\frac{B^{2}(m,z)}{2\sigma^{2}(m)}\right],
\label{eq:massfunc}
\end{eqnarray}
where $\sigma^{2}(m)$ is the variance of density fluctuations on the
scale of mass $m$, and $B$ is the critical overdensity.  If the mean
density within a region of mass $m$ is higher than $B$, then it is
ionized; $B_{0}$ is the limiting value of $B$ as $m\rightarrow\infty$.
The expression is analogous to the Press--Schechter mass function,
except that the value of the critical overdensity is different and
mass--dependent.  This formalism requires us to specify a parameter
$M_{min}$, which is the mass of the smallest collapsed halo that can
produce winds, or a hot bubble. Our fiducial value of $M_{min}$
throughout this paper is set to be $10^{11}~{\rm M_\odot}$ , but we
also consider a range of values $M_{min}=10^{9},10^{10}$ or
$10^{12}~{\rm M_\odot}$ . The choice for the lowest value is motivated
by the expectation that the cooling and collapse of gas, and therefore
star--formation in smaller halos is prevented by the UV background
(Efstathiou 1992; Thoul \& Weinberg 1996; Dijkstra et al. 2004),
whereas the highest value corresponds roughly to the largest masses
considered for LBGs at $z\sim 3$ (Somerville et al. 2002).

We assume that the temperature in these hot bubbles is sufficiently
high ($\gsim 5\times 10^{4}$K) for hydrogen to be essentially
completely ionized, and that these regions therefore produce
negligible Lyman $\alpha$ absorption in the spectra of background
quasars.  The signature of such a hot bubble intersecting a quasar
sight--line would therefore be a ``void'' in the Lyman--$\alpha$
forest. We are now in the position to compute the size distribution of
these voids; the results will be explicitly calculated and shown in
\S~\ref{sec:stat} below.

\section{Observational Data}
\label {sec:data}

\begin{figure}[t]\plotone{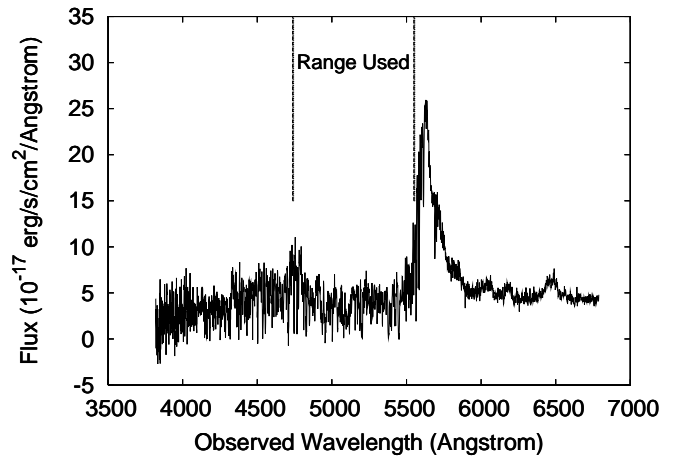}\caption{The Lyman--$\alpha$ absorption
    spectrum of a typical source used in our analysis. The two
    vertical dashed lines mark the range of the spectrum we utilized.
    This range was chosen to avoid wavelengths that contain
    Lyman--$\beta$ absorption, and the proximity region within 10 Mpc
    of the quasar.  }\label{fig:agevel}\end{figure}

In this section, we briefly describe the data we used for our
analysis.  The spectra were selected from the SDSS Data Release 4
(DR4; Adelman-McCarthy et al. 2006).  We examined 137 quasar spectra
with redshifts in the range $3.5<z<4$. We cherry--picked these
high--quality spectra by hand from 798 among the brightest quasars in
DR4 in this redshift range.  Spectra were selected so that the S/N is
greater than 8, and the wavelength resolution is about one \AA ngstrom
per pixel. From every spectrum, we only used the Lyman--$\alpha$
region from $1025(1+z)$\AA\ to $1215(1+z)$\AA, discarding shorter
wavelengths subject to additional Lyman $\beta$ absorption. In order
to avoid having to model the proximity effect, we also excised the
wavelength range corresponding to radial separations of $\leq 10$Mpc
from the source quasar. The total wavelength range we analyzed is
112203 \AA, which is equivalent to an effective redshift range of
$\Delta z=92.3$. The median redshift of the wavelength range we
utilized is $\bar z=3.3$. For illustration, in Figure~\ref{fig:agevel}
we show the spectrum of a typical source.

The raw data include the flux $f_i$ and the noise $n_i$ for each
$\sim1$\AA\ wide wavelength bin centered at $\lambda_i$. We first fit
the continuum for every spectrum, using the same clipped--variance
estimator continuum technique employed for the SDSS absorption line
catalog (e.g., York et al. 2005).  We then search the Lyman--$\alpha$
part of the spectrum for voids larger than 5\AA. We neglected smaller
voids because of the limitation from spectrum resolution, and in order
to avoid small--scale correlations between pixels induced by
large--scale structures (the size of the smallest void we consider,
$\sim 3h^{-1}$Mpc, exceeds the correlation length of the absorption
lines by a factor of $\approx$10; e.g. Cristiani et al. 1997).  Here
we define a void to be a contiguous range of neighboring pixels where
the flux--to--continuum ratio exceeds a certain threshold.  The
threshold could, in principle, be very high ( >99\%) from the simple
theoretical speculation above.  However, noise in the data limits our
choice for the threshold to be smaller than or equal to 80 percent
(see a more detailed discussion in \S~\ref{sec:stat} below).  This,
means that, in effect, we allow hot bubbles to contain some residual
neutral hydrogen causing $\sim 10\%$ absorption. We create a size
distribution of the voids, i.e. a histogram using discrete wavelength
bins of width in the range 5\AA$\leq\Delta\lambda \leq 70$\AA\, in
increments of 5\AA. We generated mock histograms using different
fitting functions, derived from models with or without hot bubbles
(see \S~\ref{sec:stat}).  The goodness of fit and the likelihood of
each model is obtained from the usual $\chi^{2}$ statistic,
\begin{eqnarray}
\chi^{2}=\sum_{i} \frac{(N_{i}-n_{i})^{2}}{n_{i}},
\end{eqnarray}
where $N_{i}$ and $n_{i}$ are the number of observed {\it vs}
predicted voids in bin $i$, respectively.

\section{Statistical Analysis}
\label{sec:stat}

In \S~\ref{sec:model}, we explained how we calculated the bubble size
distribution. In this section, we will convert this bubble size
distribution into the Lyman--$\alpha$ void distribution that could be
compared to the observational data.  In general, voids in the observed
absorption spectrum could be produced in two ways. First, in the usual
picture for the undisturbed Lyman $\alpha$ forest, due to the density
fluctuations, the IGM will contain low density regions that produce
little absorption.  Second, the presence of hot bubbles can produce
additional voids, as we explained above.  For clarity, we refer to
these two kinds of voids as "density voids" and "ionization voids",
respectively.  In the first half of this section, we discuss the
void--size distribution, including the non--trivial overlap between
individual voids, ignoring, for simplicity, the presence of noise.  In
the second half of this section, we discuss the impact of noise on our
predicted void-size distributions.

\subsection{The Expected Noise-Free Void--Size Distribution}
\label{subsec:voidsizedistribution}

The total number of voids is not simply the sum of density voids and
ionization voids, since these voids can mix in a non--trivial way.
For example, a hot bubble may be expanding into a low--density region
in the IGM, producing an ionization void that connects with an
adjacent density void, forming a single, larger apparent void.
Bearing this in mind, we first calculate the size distribution of
ionization voids.  We neglect peculiar velocities in our analysis.
Typical peculiar velocities at $z\approx 3$ on the relevant large
scales are $\lsim 100$ km/s (e.g. Gnedin \& Hamilton 2002), which
would correspond to $\sim 1.6$\AA\ shifts in the apparent
spectrum. This is a small fraction of the smallest void size we
consider. Furthermore, peculiar velocities are proportional to the
overdensity and will be smaller for spectral pixels of interest that
have lower-than-usual absorption.  In this case, the size of an
ionization void is determined solely by the Hubble flow, which in turn
scales directly with the size of the hot bubbles. For simplicity, and
consistent with the model assumptions in the previous section, we
further assume that the bubbles are spherical. For a given mass $m$,
the volume of a bubble is then
\begin{eqnarray}
V=\frac{m}{\bar{\rho}(1+\delta_{m})}, \label{eq:volume}
\end{eqnarray}
where $\bar{\rho}$ is the mean density of the universe, and
$\delta_{m}$ is the overdensity within the bubble. By construction, in
the hot bubbles, $\delta_{m}$ is equal to $B$ in equation
(\ref{eq:massfunc}). Let as assume that the line of sight (LOS)
intersects a bubble of radius $R=R(m)$ at an impact parameter $0\leq
b\leq R$ (defined as the distance of closest approach between the
center of the hot bubble and the LOS).  The length of LOS within the
hot bubble is then $l=2\sqrt{R^{2}-b^{2}}$, and the Hubble velocity
across this region is
\begin{eqnarray}
v_{H}=H(z)l=2H(z)\sqrt{R^{2}-b^{2}} \label{eq:hubbleflow}.
\end{eqnarray}
The hot region produces a void covering the range of observed wavelengths
\begin{eqnarray}
\Delta\lambda=\lambda_\alpha(1+z) \frac{v_{H}}{c}\equiv K\sqrt{R^2-b^2},
\label{eq:gapsize}
\end{eqnarray}
where $\lambda_\alpha=1215$\AA\ is the rest-frame Lyman--$\alpha$
wavelength, $c$ is the speed of light, and in the last step we have
defined $K\equiv 2\lambda_\alpha(1+z)c^{-1}H(z)$. The number density of
voids of size $\Delta\lambda$ along a given LOS per unit redshift and
unit size is obtained directly from the mass function through the
equation
\begin{eqnarray}
\nonumber
\frac{\mathrm{d}^{2}N}{\mathrm{d}\Delta\lambda\mathrm{d}z}(z,\Delta\lambda) & = &
\int_{m_{\rm min}}^{\infty}\mathrm{d}m
\int_{0}^{R}\mathrm{d}b\, 2\pi b\,
\delta(\Delta\lambda-K\sqrt{R^2-b^2})\times\\
&& \times \frac{\mathrm{d}n}{\mathrm{d}m}
\frac{c(1+z)^2}{H(z)}.
\label{eq:dn_dz0}
\end{eqnarray}
In this equation, $dn/dm$ is the mass function of hot bubbles, $m_{\rm
min}=m_{\rm min}(\Delta\lambda)$ is the smallest hot bubble that can
produce a void of length $\Delta\lambda$ (at $b=0$),
$d^2V/dzd\Omega=cH^{-1}(z)(1+z)^2d_A^2(z)$ is the comoving volume per
unit redshift and solid angle, and $2\pi b \mathrm{d}b d_A^{-2}(z)$ is
the solid angle extended by a narrow circular annulus at impact
parameter $b$ and width $\mathrm{d}b$.  Note that the angular diameter
distance, $d_A(z)$, drops out of the equation. The Dirac delta
function in the top row enforces the relation between bubble radius
$R$ and impact parameter $b$ to produce a void of fixed length
$\Delta\lambda$.  Performing the $b$--integral (using the property of
the delta function $\delta(f(x))=\delta(x)/|f^\prime|$), we find
\begin{eqnarray}
\frac{\mathrm{d}^{2}N}{\mathrm{d}\Delta\lambda
\mathrm{d}z}(z,\Delta\lambda) = 
\frac{\pi c^{3}}{2 \lambda_\alpha^{2}}
\frac{\Delta\lambda}{H^{3}(z)}
\int_{m_{\rm min}}^{\infty}\mathrm{d}m\frac{\mathrm{d}n}{\mathrm{d}m}
\label{eq:dn_dz}
\end{eqnarray}

To calculate the overall distribution of voids, we need to take into
account both density voids and ionization voids, and the list of
possible overlaps between them.  If there were only density voids,
their distribution would obey a simple exponential.  This follows
directly from the Poisson distribution of absorption lines (e.g. 
Crotts 1987; Ostriker et al. 1988) and assumes that 
the spatial correlations between different absorption lines on the
large scales of interest ($\gsim 10$ times the correlation
length of absorption lines; e.g. Cristiani et al. 1997) are negligible.
The number of pure density voids of size $\Delta\lambda$, per unit
observed total wavelength range $\Delta\lambda_{tot}$, and per unit
void size $\Delta\lambda$, is then given by
\begin{eqnarray}
\frac{\mathrm{d}^2N_{0}}{\mathrm{d}\Delta\lambda\mathrm{d}\Delta\lambda_{tot}} = A \exp(-b\Delta\lambda) \label{eq:p0}
\end{eqnarray}
Here $b$ is related to the mean number of absorption lines per unit
wavelength, above a given threshold strength, weighted appropriately
over redshift (see below).  $A$ is a normalization factor which also
depends on the number density of absorption lines and also, in general
on $\Delta\lambda_{tot}$.  Note that $A$ has units of
(wavelength)$^{-2}$. The simplest case is that of a single redshift,
negligible noise, and absorption lines that do not blend together (we
will discuss the issue of noise further below). In this case, one can
compute the total number of expected voids, and uniquely compute the
normalization $A$ as a function of $b$ and $\Delta\lambda_{tot}$;
$A=A(b,\Delta\lambda_{tot})$. Taking the limiting case of
$\Delta\lambda_{tot}\rightarrow\infty$, and the width of individual
absorption lines approaching zero, we find $A\rightarrow b^2$.  This
unique correspondance, however, is spoiled when the above assumptions
are relaxed. In order to avoid addressing these issues, or having to
model the mean transmission of the forest and its evolution with
redshift self--consistently, we treat $A$ as an independent free
parameter in our fitting procedure.  It is worth noting, however, that
further modeling could significantly tighten the constraints we derive
below on the abundance of hot bubbles. In practice, we find deviations
of up to a factor of $\sim 2$ from the above limiting formula $A=b^2$,
implying that the above simplifying assumptions do not introduce gross
errors.

The presence of ionization voids will disrupt the exponential
distribution of the density voids.  Let us first consider an observed
void of size $\Delta\lambda$ that contains exactly one ionization void
of some size $s\leq\Delta\lambda$, overlapping with zero, one, or two
neighboring density voids whose sizes add up to $\Delta\lambda-s$. The
number of such voids, per unit observed total wavelength range
$\Delta\lambda_{tot}$, and per unit void size $\Delta\lambda$, is
given at some redshift $z$ by
\begin{eqnarray}
\frac{\mathrm{d}^2N_{1}}{\mathrm{d}\Delta\lambda\mathrm{d}\Delta\lambda_{tot}}
&=& \frac{A}{\lambda_\alpha}
\int_{0}^{\Delta\lambda}\mathrm{d}s \frac{\mathrm{d}^2N}{\mathrm{d}\Delta\lambda
\mathrm{d}z}(s)\nonumber\\
&& \times \exp[-b(\Delta\lambda-s)](\Delta\lambda-s),
\label{eq:p1}
\end{eqnarray}
The equation follows from noting that (i) the center of
the ionization void of size $s$ can be placed anywhere over an
interval of length ($\Delta\lambda-s$), and (ii) none of the pixels in
the remaining length $(\Delta\lambda$--$s)$ that are not covered by
the ionization void should have an absorption line.  Note that in the
limit of $b\rightarrow\infty$ and $A\rightarrow b^2$, density voids
will be rare, and equation~(\ref{eq:p1}) indeed reduces to the
abundance of ionization voids (eq. \ref{eq:dn_dz}), as it should.

Similarly, the observed void of size $\Delta\lambda$ could contain two
ionization voids, of sizes $(s,u)\leq\Delta\lambda$, connecting with
neighboring density voids. The number of these cases is given by an
argument analogous to the previous case, except we need to enforce the
condition that the two ionization bubbles are, by definition,
disjoint.  This can be achieved by the following procedure: (i) choose
a size $0<s<\Delta\lambda$ for the first ionization void, (ii) then
choose a location for this void, measured by the distance $s\leq t <
\Delta\lambda$ of the right ``edge'' of the $s$ void from the left
``edge'' of the larger $\Delta\lambda$ void, and (iii) finally place a
second ionization void, of size $0<u\leq (\Delta\lambda-t)$ anywhere
in the remaining interval $(\Delta\lambda-u-t)$.  We find, accordingly,
\begin{eqnarray}
\frac{\mathrm{d}^2N_{2}}{\mathrm{d}\Delta\lambda\mathrm{d}\Delta\lambda_{tot}}
&=&\frac{A}{\lambda_\alpha^2} 
\int_{0}^{\Delta\lambda} \mathrm{d}s
\int_{s}^{\Delta\lambda} \mathrm{d}t
\int_{0}^{\Delta\lambda-t} \mathrm{d}u
\frac{\mathrm{d}^2N}{\mathrm{d}\Delta\lambda
\mathrm{d}z}(s)\frac{\mathrm{d}^2N}{\mathrm{d}\Delta\lambda
\mathrm{d}z}(u)\nonumber\\ & &\times
\exp[-b(\Delta\lambda-s-u)](\Delta\lambda-u-t)
\label{eq:p2}
\end{eqnarray}

The total number of voids of size $\Delta\lambda$ is then given by 
\begin{eqnarray}
\frac{\mathrm{d}^2N}{\mathrm{d}\Delta\lambda\mathrm{d}\Delta\lambda_{tot}}
&=&
\frac{\mathrm{d}^2N_{0}}{\mathrm{d}\Delta\lambda\mathrm{d}\Delta\lambda_{tot}}+
\frac{\mathrm{d}^2N_{1}}{\mathrm{d}\Delta\lambda\mathrm{d}\Delta\lambda_{tot}}+
\frac{\mathrm{d}^2N_{2}}{\mathrm{d}\Delta\lambda\mathrm{d}\Delta\lambda_{tot}}+... \nonumber\\
n&=&n_{0}+n_{1}+n_{2}+...,
\label{eq:ptotal}
\end{eqnarray}
where we introduced $n = d^2N/d\Delta\lambda d\Delta\lambda_{tot}$ to
simplify notation. In the rest of this paper, we omit terms in the
above formula above ``second order'' (the two ionization void case);
we will discuss the effect of higher order terms in
\S~\ref{sec:discussion} below.

As a first check, to see whether hot bubbles have a conspicuous effect
on the Lyman--$\alpha$ spectrum, we fit the observed histogram for
$n(\Delta\lambda$) using the exponential function $n_0$ alone.  The
exponential fit turns out to be adequate, immediately revealing that
we cannot rule out the null hypothesis that the data contains no hot
bubbles. Then we used our model formula in equation~(\ref{eq:ptotal})
to fit the data. For a given $\zeta$, we adjusted our free parameters
$A$ and $b$ to minimize $\chi^2$ and obtain the maximum
probability. We start with $\zeta=1$ and increase $\zeta$ gradually in
increments of 0.01 until the fit breaks down, i.e. until the maximum
probability is smaller than a threshold value. We chose the fiducial
value of $10^{-3}$ for this threshold probability in our analysis (see
discussion below).

The procedure outlined above yields an upper limit of $\zeta$ that
corresponds to the threshold probability. Equivalently, we can convert
$\zeta$ to a corresponding upper limit on the global volume--filling
factor $Q$ of the hot bubbles, using the equation

\begin{eqnarray}
Q=\int_{\zeta M_{min}}^{\infty}\frac{\mathrm{d}n}{\mathrm{d}m}\frac{m}{\bar{\rho}(1+B(m,z))}\mathrm{d}m.
\label{eq:filfactor}
\end{eqnarray}

\begin{figure}[t]\plotone{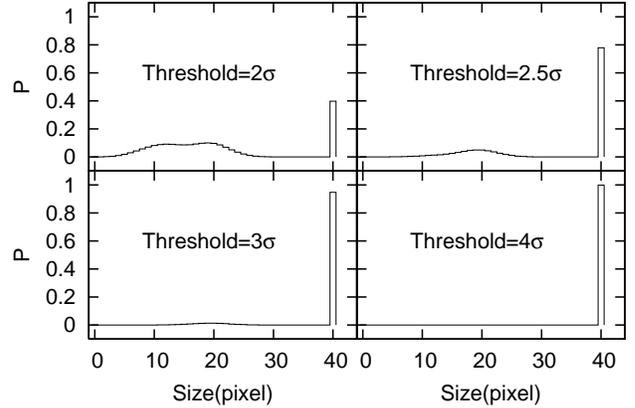}
  \caption{The figure illustrates the impact of noise on the
    void--size distribution.  We consider a single true void whose
    size is 40 pixels, with no true absorption in any of these
    pixels. Fake absorption spikes, due to Gaussian noise, are then
    added.  The noise spikes sub--divide the 40--pixel void into many
    smaller sub--voids, producing a distribution of void--sizes that
    depends on the adopted threshold for defining a void.  The four
    panels show the resulting sub--void size distributions, with four
    different choices for the threshold. The choices correspond to
    decrements below the continuum set at $2$, $2.5$, $3$, and $4$
    times the noise. In the latter two cases (bottom two panels) the
    distribution approaches a delta function at the 40-pixel size,
    showing that noise has little effect when the threshold is smaller
    than $\sim 2.5\sigma$ (or about $75\%$ of the continuum level for
    the $S/N\approx 10$ spectra we used in our
    analysis).}\label{fig:voidsplit}\end{figure}

\subsection{The Impact of Spectral Noise }
\label{subsec:noise}

Before describing our results, we discuss the choice of our absorption
threshold for defining a void in the data, and the impact of noise in
the spectra. The average S/N in the data we utilize is S/N=9.94,
which, essentially, forced us to select a low threshold in our
definition of a void. Note that the relatively lenient observational
threshold means that some residual absorption is allowed to take place
in the hot bubbles. In our simple treatment, Lyman--$\alpha$ forest
absorption lines are randomly distributed, yielding the exponential
distribution of the density voids. However, spectral noise tends to
produce some additional, fake absorption lines, which will break up
true voids.  Fortunately, noise at different pixels can be treated as
uncorrelated over scales of more than a few pixels.  If S/N were a
constant over the whole wavelength range, the exponential void
distribution would remain valid -- the effect of the noise could be
absorbed in the uninteresting (for us) constants $A$ and $b$.

Unfortunately, in practice, the S/N is usually larger in pixels where
the flux is larger, which, in general, would necessitate further
modeling.  However, this complication can be avoided by choosing a
sufficiently low threshold in defining a void, such that the number of
fake absorption lines in the wavelength range we analyze is small.
For a rough illustration, let us assume that the noise is Gaussian,
with 1$\sigma$ values corresponding to $\approx 10\%$ of the
unabsorbed flux. Since we use $\approx 10^5$ independent pixels in our
analysis, we expect roughly 2250, 600, 135, 25, and 3 pixels with
absorption lines extending below thresholds of 80, 75, 70, 65, and 60
percent of the continuum.  As we shall find below, there are 3632,
4064, 4345, 4573, and 4667 voids in these cases, occupying a total
number of 37545, 44900, 51558, 58303, and 64833 pixels.  Therefore,
the fractional increase in the number of voids in these cases,
assuming each fake absorption line results in one extra void, is
(2250/112203)(37545/3632)=0.21, etc.  Clearly, this fractional
increase is small for thresholds below 70 percent.

Next, let us consider including the effect of noise explicitly in our
model fitting function (\ref{eq:ptotal}). Rather than modeling the
noise in the data, we added noise into our theoretical void
distribution, before comparing it with the data. The effect of
uncorrelated constant S/N noise on the density void distribution $n_0$
is automatically absorbed into the free parameters $A$ and $b$
(i.e. the noise changes only the values of $A$ and $b$, which are
anyway free parameters in our analysis). Including the effect of noise
on the ionization void distribution is more subtle. First, we assume
that there is no absorption in the ionization voids, and the effect of
a noise absorption--spike is to cut a large ionization void into two
smaller ones. This amounts to a re--distribution of
$\frac{\mathrm{d}^{2}N}{\mathrm{d}\Delta\lambda \mathrm{d}z}$. Let us
consider an ionization void containing $x$ pixels (since each pixel is
$\sim 1$\AA, this also roughly gives the size of the void in \AA). The
probability that a void of size $y \leqslant x$ appears, as a result
of noise sub--dividing the true void of size $x$ (with $x$ and $y$ 
regarded as the integer number of pixels) is
\begin{eqnarray}
p(y|x)&=& \sum_{i=0}^{x-y} {x\choose i} p^{i} (1-p)^{x-i} \nonumber\\
& &\times 
{x-i \choose y}
\left(\frac{1}{1+i}\right)^{y}\left(\frac{i}{1+i}\right)^{x-y-i}(1+i)\label{eq:noiseadded}
\end{eqnarray}
Here, $p$ is the probability that the noise in a given single pixel
lowers the flux down below the threshold. For instance, if the
threshold is chosen at 90\% of the continuum, or $1\sigma$, then we
have $p=0.16$. The first line in the summation on the right hand side
of equation~(\ref{eq:noiseadded}), ${x\choose i} p^{i} (1-p)^{x-i}$,
gives the probability that exactly $i$ pixels, selected randomly from
among $x$ pixels, are lowered below the threshold.  These $i$ pixels
then divide the whole void into $i+1$ smaller sub--voids (allowing the
length of a sub--void to be zero, in cases when noise spikes occupy
neighboring pixels).  The term ${x-i\choose
y}(\frac{1}{1+i})^{y}(\frac{i}{1+i})^{x-y-i}$ gives the probability
that a given sub--void's length is exactly $y$.  Finally, we multiply
this last factor by the number of sub--voids, $(i+1)$, and sum over
all the possible $i$'s.

As an illustration of the impact of noise, in
Figure~\ref{fig:voidsplit} we consider a 40--pixel void, which is
allowed to be sub--divided by noise. The four different panels
correspond to different choices for the threshold to define a void. In
the absence of noise, we would have a single 40-pixel void in each
panel. The effect of noise is to produce a new distribution of smaller
voids, which can be regarded as an asymmetric kernel, by which the
actual noise--free distribution $n$ should be convolved. As the figure
shows, when the threshold is chosen to be lower than $2.5\sigma$ (or
at $\sim 75\%$ of the continuum in our case), the noise has little
effect on the void size distribution.  Finally, we note a complication
that arises during void mixing. When noise is ignored, any ionization
void can connect with density voids on both sides. But if an
ionization void divided, the sub--voids can only connect on one side
(if the sub-void is on the edge of the original void) or neither side
(if the sub-void is flanked on both sides by noise spikes). In our
numerical calculation of the noisy void-size distribution, we kept
track of these different types of voids, and treated them
accordingly. This entails modifying equations~ (\ref{eq:p1}) and
(\ref{eq:p2}), which describe only the case when ionization voids
connect on both sides with density voids; in the interest of brevity,
we do not list these modified equations here.

\section{Results}
\label{sec:results}

\begin{table}[ct]
  \caption{Best fitting exponential distributions of the form
    $A\exp[-b\Delta\lambda]$ (i.e. models of the Lyman $\alpha$ forest
    without any hot bubbles), to the observed histogram of void
    sizes. Five different thresholds are considered for defining
    voids. }
    \label{tbl:expfit}
\begin{center}
\begin{tabular}{|c|c|c|c|c|}
\hline \hline
Threshold  & A & b & $\chi^{2}$ & Likelihood \\
\hline
\hline 80\% & 0.0192 & 0.198 & 1.093 & 0.198\\
\hline 75\% & 0.0164 & 0.173 & 0.437 & 0.823\\
\hline 70\% & 0.0136 & 0.150 & 0.529 & 0.757\\
\hline 65\% & 0.0114 & 0.132 & 0.812 & 0.481\\
\hline 60\% & 0.0092 & 0.114 & 0.762 & 0.539\\
 \hline \hline
\end{tabular}
\end{center}
\end{table}

\begin{table}[ct]
  \caption{Upper limits on $\zeta$ and on the corresponding volume
    filling factor $Q$ of hot bubbles, in the model for the Lyman
    $\alpha$ forest that includes such hot bubbles. The fit is
    considered unacceptable when the likelihood drops below
    $10^{-3}$. For comparison, we also calculated the upper limits of
    $\zeta$ and $Q$ with less stringent likelihood thresholds of
    $10^{-2}$, and also when the formula in equation~(\ref{eq:ptotal})
    is cut at the first order (the probability threshold is
    $10^{-3}$). These results are listed in the first and second
    columns within parentheses, respectively.}
\label{tbl:newfit}
\begin{center}
\begin{tabular}{|c|c|c|}
\hline \hline
Threshold & $\zeta$ & Volume filling factor $Q$ \\
\hline
\hline 80\% & 5.70 (5.64,5.72)& 0.226 (0.221,0.227)\\
\hline 75\% & 5.14 (4.94,5.44)& 0.185 (0.172,0.206)\\
\hline 70\% & 4.89 (4.69,5.31)& 0.169 (0.156,0.197)\\
\hline 65\% & 4.94 (4.72,5.42)& 0.172 (0.158,0.205)\\
\hline 60\% & 5.77 (5.61,6.35)& 0.231 (0.219,0.281)\\
 \hline \hline
\end{tabular}
\end{center}
\end{table}

We first list, in Table 1, the result of the exponential fit. As we
can see, the $\chi^2$ likelihoods in this Table, for all five choices
of the threshold, are $\gsim 20$ percent. This means that the
exponential fits are acceptable, and there is no statistical evidence
for hot bubbles, or for any voids beyond those found in the usual
``fluctuating Gunn--Peterson absorption'' picture for the Lyman
$\alpha$ forest.

Next, we constrain the abundance of hot bubbles using our fitting
formula in equation~(\ref{eq:ptotal}), modified to include the effects
of a constant Gaussian noise as discussed in the preceding
section. Table 2 gives the maximum values of $\zeta$, and the
corresponding maximum allowed hot bubble volume filling factors ($Q$),
with $\chi^2$ likelihoods at $10^{-3}$.  The tightest constraint we
find is for a threshold of $70\%$, in which case the volume filling
factor of hot bubbles is at most 16.9\%.  Interestingly, the constraint
does not vary monotonically. As the threshold is lowered below 70\%,
the number of large voids obtained from the spectra increases, which
weakens our constraints.  As the threshold is increased above 70\%,
our constraint again becomes weaker, because spectral noise in this
case can eliminate the large voids that the models would otherwise
predict.

\begin{figure}
  \begin{center}
    \begin{tabular}{cc}\rotatebox{-0}{
      \resizebox{80mm}{!}{\includegraphics{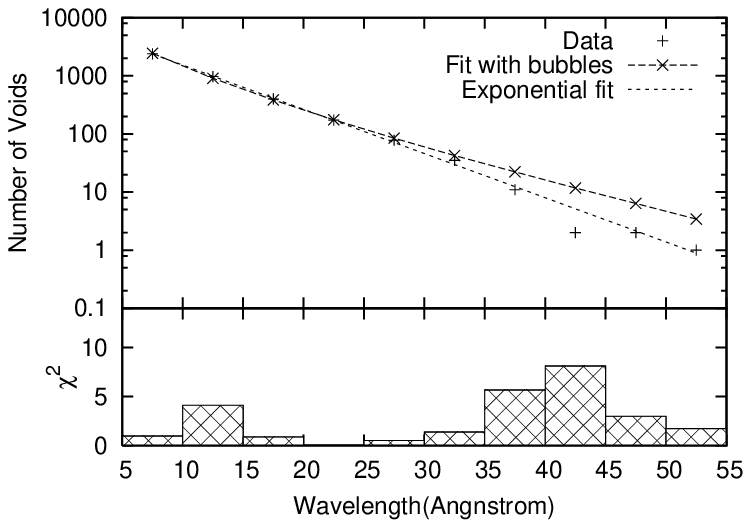}}}\\
    \rotatebox{-0}{ \resizebox{80mm}{!}{\includegraphics{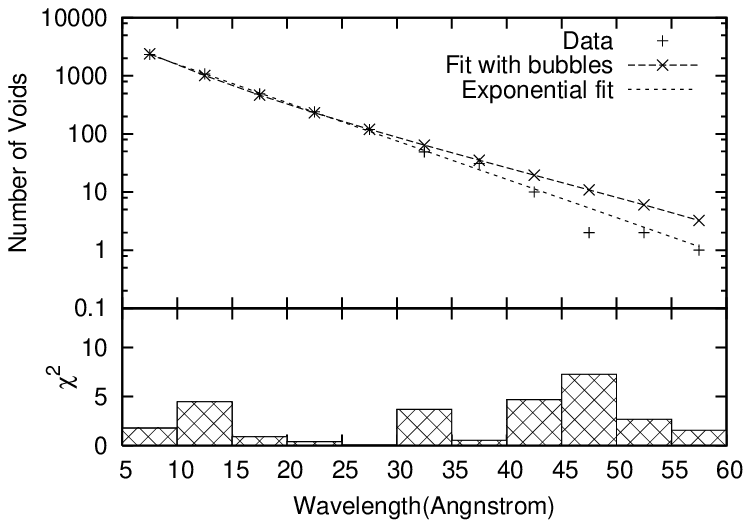}}}
    \end{tabular}
    \caption{Results of fitting model void size histograms to the
      data. The threshold for defining a void is 75\% (70\%) of the
      unabsorbed continuum in the upper (lower) panel. The crosses
      show the data points inferred from 137 quasar spectra; the
      dashed curves show the best fitting results of our fitting
      formula (\ref{eq:ptotal}) when $\zeta$ is at its upper limit
      (i.e. with the largest allowed filling factor of hot bubbles),
      while the dotted curves correspond to the exponential fits
      (i.e. models without hot bubbles). We also show the contribution
      from each void-size bin to the total $\chi^{2}$.  Models with
      large filling factors can be ruled out primarily because they
      over--produce the number of $40-50$\AA\ voids; they also
      under--produce the number of $10-15$\AA\ voids.}
    \label{fig:fitting}
  \end{center}
\end{figure}

In order to investigate the importance of the likelihood threshold, in
Table 2 we list within parentheses (first column) the upper limit on
$\zeta$ when the probability threshold is increased to $10^{-2}$. As
we can see, the results change only a little.  This is not surprising:
we find that probabilities fall very rapidly with increasing $\zeta$
once they reach a level below $10^{-2}$. In Figure~\ref{fig:fitting},
we show two explicit void--size histograms used in our fitting
procedure. The threshold is chosen to be 70\% in the lower panel, and
75\% in the upper panel; the $\zeta$'s are at their upper limits
listed in Table 2. In the same figure, we also plot the contribution
from each void-size bin to the total $\chi^{2}$. As we can see, the
hot bubbles not only increase the number of large voids in the
Lyman--$\alpha$ forest, but also disturb the distribution of smaller
voids.  In other words, the primary reason we can rule out these
models with large bubble filling factors is that they overproduce the
number of $40-50$\AA\ voids; however, they also significantly
underproduce the number of smaller, $10-15$\AA\ voids. We find that
the free choice of $A$ always assures a good fit to the
size--distribution of the intermediate size voids, in the range
$15-40$\AA.

To investigate the importance of the assumed value of $M_{min}$, in
the range motivated in the Introduction, we replace our fiducial value
of $M_{min}=10^{11}{\rm M_\odot}$ by $10^{9}$, $10^{10}$ and $10^{12}
{\rm M_\odot}$, and repeat our analysis in each case.  The results are
listed in Table \ref{tbl:minmass}. The upper limits of $\zeta$
generally increase as $M_{min}$ increases. Physically, this is because
a single collapsed clump is allowed to ionize a larger region, when
the total number of such ionizing clumps is reduced, due to an
increase in $M_{min}$. The volume filling factors, on the other hand,
decrease with increasing $M_{min}$. This is because low--mass halos
tend to produce small ionized regions, which are too small to
effectively disturb the exponential void distribution at sizes above
$5$\AA, but these small bubbles still contribute to the volume filling
factor. The increase in $M_{min}$ eliminates these smaller ionized
regions, and such models are therefore easier to constrain. For
example, for $M_{min}=10^{12}{\rm M_\odot}$, we find a relatively
tight limit of $Q\lsim 11\%$.

\begin{table*}[ht]
  \caption{Upper limits on $\zeta$ and on the volume filling factors
    of hot bubbles, when the minimum halo mass to produce a bubble,
    $M_{min}$, is assumed to be $10^{9}$, $10^{10}$, $10^{11}$ and
    $10^{12}{\rm M_\odot}$.}
\label{tbl:minmass}
\begin{center}
\begin{tabular}{|c||c|c||c|c||c|c||c|c|}
\hline \hline
 $M_{min}$& \multicolumn{2}{c||}{$10^{9}{\rm M_\odot}$}& \multicolumn{2}{c||}{$10^{10}{\rm M_\odot}$}&\multicolumn{2}{c||}{$10^{11}{\rm M_\odot}$}&\multicolumn{2}{c|}{$10^{12}{\rm M_\odot}$}\\
\hline
Threshold & $\zeta$ & $Q$& $\zeta$ & $Q$ & $\zeta$ & $Q$ &$\zeta$ & $Q$  \\
\hline
\hline 80\% & 2.39 & 0.314 & 3.28 & 0.278 & 5.70 & 0.226 & 16.50 & 0.146\\
\hline 75\% & 2.27 & 0.271 & 3.05 & 0.234 & 5.14 & 0.185 & 14.55 & 0.118\\
\hline 70\% & 2.20 & 0.249 & 2.94 & 0.215 & 4.89 & 0.169 & 13.76 & 0.108\\
\hline 65\% & 2.21 & 0.252 & 2.95 & 0.217 & 4.94 & 0.172 & 13.97 & 0.111\\
\hline 60\% & 2.38 & 0.310 & 3.27 & 0.276 & 5.77 & 0.231 & 18.06 & 0.170\\

 \hline \hline

\end{tabular}
\end{center}
\end{table*}

\section{Discussion}
\label{sec:discussion}

It is interesting to ask whether the constraints we obtained above,
using 137 quasars, could become significantly tighter by simply
increasing the number of the spectra.  The SDSS data--base (up to DR4)
contains approximately 30,000 quasar spectra at redshifts $z>2.3$.  At
a fixed value of $A$ and $b$, $\chi^{2}$ will increase roughly in
proportion to the number of quasar spectra.  To quantify the effect of
this increase on the upper limit on $Q$, we generated mock void--size
histograms implementing realizations of the exact exponential void
distribution with $A$ and $b=$ chosen to be the best--fit values from
Table 1.  As a test of the method, we first generated histograms
corresponding to 137 quasar spectra. When fitting our model to these
mock data, we found an upper limit $Q<16.9\%$ at the threshold of
70\%, in agreement with the results in Table~2 using the actual
spectra.  Next, we generated mock histograms for a hypothetical 13,700
quasar spectra.  We found that this 100--fold increase in the number
of quasars improved the upper limit on the volume filling factor to
$Q<6.6\%$.

Another way to improve the sensitivity of the constraints would be to
use spectra with higher signal-to-noise ratio and/or with higher
resolution.  To illustrate the impact of noise, we assumed noise is
negligible, and repeated our analysis for the 80\% threshold case. We
found that the upper limits on $\zeta$ and $Q$ improve from (5.70,
0.226) to (4.67,0.155), respectively.  We expect the limits to tighten
further if we raise our threshold, as would be possible if the
spectral noise was indeed very small.
We expect the main improvement allowed by higher resolution spectra is
that we could utilize the void--size distribution down to smaller
sizes, below 5\AA.  This would be significantly more than just an
additional bin of data. This is because as $\zeta$ is decreased below
some value, the typical bubble radius produced around a single
collapsed halo will become smaller than the mean distance between
collapsed halos.  In this case, there will be very little overlap
between different hot bubbles, and the ionization voids will become
small.  Constraints on $\zeta$ and the filling factor of such small
voids would only be possible in higher resolution spectra.  The
correlation between Lyman--$\alpha$ absorption lines can no longer be
ignored on scales smaller than we utilized here, and would have to be
modeled in analyzing higher--resolution spectra.

The main virtue of our model is that it takes bubble mergers into
account.  Nevertheless, it is based on assumptions that are likely to
be oversimplifications.  First, we treated $\zeta$ as a constant,
while it is possible that it could be strongly dependent on the mass
and the environment of the collapsed halo, and could also evolve with
redshift.  For example, the star--formation rate may scale roughly
linearly with the mass of the gas reservoir, and hence with the halo
mass. However, the gravitational binding energy per unit mass in a
halo of mass $M$ scales as $kT\propto M^{2/3}(1+z)$, making it more
difficult for winds to escape from larger halos.  Furthermore, in its
original context of reionization, the bubble--merger model we adopted
here was motivated by the reasonable assumption that the merger of two
photoionized bubbles conserves the total ionized mass.  If the hot
bubbles are produced by overlapping galactic winds (rather than
photoionization), then this assumption is likely to be much less
accurate. When two winds overlap, they will interact dynamically, and
the winds will not instantanously propagate to the edge of the joint
bubble, to conserve mass.  As a result, it is likely that the
effective value of $\zeta$ will further decrease as the overlap
between winds becomes more significant.  It would be possible to
incorporate an $M$ and $z$--dependence, $\zeta=\zeta(M,z)$, in our
modeling, as well as a further decrease in $\zeta$ that depends on the
number of galaxies per bubble, but we leave such improvements to
future work.

Our modeling also assumes spherical ``hollow'' bubbles, and ignores
their inner structures. It is possible for bubbles to be quite
non--spherical; this would be similar to smoothing the
size--distribution with an appropriate scatter, representing the
dispersion in radial extent when the line of sight crosses a bubble in
different directions.  The impact of such a scatter would be to
increase the number of large voids. Since the upper limit we found is
driven by the predicted number of such large voids
(Fig.~\ref{fig:fitting}), these upper limits should be improved if the
scatter due to non-sphericity was included. On the other hand, it is
also possible that there is residual neutral hydrogen within the hot
bubbles, so the voids are not completely empty, either due to
incomplete mixing between hot wind material and the ambient IGM, or
due to radiative transfer effects (if bubble heating is due to
photo--ionization).  Some numerical simulations also indicate that the
galactic winds tend to expand preferentially towards lower density
regions and leave the relatively more overdense filaments, which
produce the deeper Lyman $\alpha$ absorption lines, intact
(e.g. Theuns et al. 2002; Bruscoli et al. 2003; McDonald et al. 2005).
In particular, Theuns et al. (2002) explicitly show that in their
models for galactic winds, the winds produce no discernible effect on
the column density distribution of absorption lines even down to
column densities below $10^{12}~{\rm cm^{-2}}$.  In this case, the
winds would produce very few, if any, new voids in Lyman $\alpha$
spectra, even at stringent thresholds. This conclusion, however, may
not be generic - it must depend on the spatial distribution of sources
and the nature and geometry of winds, as well as on the filling factor
of winds (e.g. we expect winds to ultimately penetrate the denser
regions, if their filling factor is high).

Our analysis also neglects the correlations both among Lyman--$\alpha$
lines and also between Lyman--$\alpha$ lines and hot bubbles. Both
deep Lyman--$\alpha$ absorption lines and hot bubbles are inclined to
appear at overdensity regions, so there should be a positive
correlation between these two, which should be taken into account in
future modeling.  Finally, in our fitting procedure, we omit terms
above second order in equation~(\ref{eq:ptotal}). Intuitively, one
expects that higher order terms tend to produce even larger voids,
and, as Figure~\ref{fig:fitting} shows, it is these large predicted
voids that yield our constraints.  This expectation is borne out in
Table~\ref{tbl:newfit}, where we list the upper limits on $\zeta$ when
only the first order term is retained (i.e. we use $n_0+n_1$ in
equation~\ref{eq:ptotal}).  The Table shows that omitting the second
order terms weaken the constraints.

Finally, we note that the three--year data from WMAP favors a lower
value for the power spectrum normalization than the fiducial
$\sigma_8=0.9$ we adopted. We have explicitly verified, however, that
this choice has no significant effect on our conclusions.  In
particular, we repeated the calculations from Table 2, with all
parameters left unchanged, except replacing $\sigma_8=0.9$ by
$\sigma_8=0.75$. We found that this changes the upper limits on the
filling factor $Q$ by less than 3 percent -- although the
corresponding values of the efficiency $\zeta$ are increased by a
factor of $\sim$ two.  This latter change is to be expected - the
reduction in the underlying dark matter halo abundance implies that
producing the same filling factor requires a higher efficiency.

\section{Conclusions}
\label{sec:conclude}

Motivated by the empirical evidence for significant ``pre--heating''
of at least parts of the IGM at $z\sim 3$, we made a simple model for
the spatial distribution of preheated regions.  The model assumes
spherical hot bubbles around collapsed dark matter halos, and allows
these spheres to merge into larger ``super--bubbles''.  We predicted
the number of voids that such hot bubbles would produce in Lyman
$\alpha$ absorption spectra of background quasars.

Our comparison with the observed spectra of 137 quasars did not
uncover evidence for hot bubbles at $z\sim 3$.  Instead, we found an
upper limit on the volume--filling factor of hot bubbles, ranging from
11-25\%, depending on the assumed size of the smallest halo that
produces hot bubbles.  This is comparable to the the fraction of the
total mass in the present--day universe in low--mass clusters and
groups, suggesting that the pre--heating at $z\sim 3$ may not have
affected all the gas currently residing in these objects.

These constraints are complementary to studies of the impact of
galactic winds on Lyman $\alpha$ absorption spectra in the vicinity of
known galaxies (LBGs).  The latter approach is a more sensitive probe
of the effects of the LBGs themselves, whereas searching the
``global'' statistics is sensitive to feedback from undetected
galaxies whose spatial distribution is not strongly correlated with
LBGs.

While the constraints we obtain are still relatively weak, they
suggest that pre--heating, if it occurred, avoided heating the
low--density gas in the proto--cluster regions, either by operating
relatively recently ($z\lsim 3$) or by depositing entropy
preferentially in the high--density regions.   We expect that
our constraints could be improved significantly by analysing a
larger number of quasars spectra, and by improving on the simple
model presented here.

\vspace{-0.5\baselineskip}

\acknowledgments CS thanks Shouhao Zhou for useful discussions.  We
also thank Greg Bryan and Steve Furlanetto for helpful comments, and
Anne Abramson, who completed a senior thesis at Columbia University in
the Spring of 2004 on the subject of this paper, for some preliminary
work.  ZH acknowledges partial support by NSF grant AST-05-07161, by
the Initiatives in Science and Engineering (ISE) program at Columbia
University, and by NASA through grant NNG04GI88G.  AC acknowledges
partial support by the Space Telescope Science Institute from grant
AR-9195.


\end{document}